\documentclass[aps, pra, a4paper, amssymb, reprint, showpacs, nofootinbib]{revtex4-1}

\usepackage{amssymb}
\usepackage{graphicx}
\usepackage{amsmath}
\usepackage{hyperref}

\usepackage{bbold}
\usepackage{color}

\newcommand{\abs}[1]{\left|#1\right|}
\newcommand{\sabs}[1]{|#1|}

\newcommand{\smean}[1]{\langle #1 \rangle}

\newcommand{\mE}{\mathcal{E}}
\newcommand{\mA}{\mathcal{A}}

\newcommand{\mF}{\mathcal{F}}

\newcommand{\mfr}{\mathfrak{r}}
\newcommand{\mft}{\mathfrak{t}}

\newcommand{\rmd}{\textrm{d}}

\newcommand{\rmat}{\textrm{at}}
\newcommand{\rmm}{\textrm{m}}
\newcommand{\rmL}{\textrm{L}}

\newcommand{\om}{\omega}

\newcommand{\figref}[1]{Fig.\,\ref{#1}}
\newcommand{\eeqref}[1]{Eq.\,\eqref{#1}}
\newcommand{\eqsref}[1]{Eqs.\,\eqref{#1}}
\newcommand{\secref}[1]{Sec.\,\ref{#1}}
\newcommand{\appref}[1]{Appendix \ref{#1}}

\begin{document}
\date{\today}

\title{Cavity-Enhanced Long-Distance Coupling of an Atomic Ensemble to a Micromechanical Membrane}

\author{B. Vogell}
\author{K. Stannigel}
\author{P. Zoller}
\affiliation{Institute for Theoretical Physics, University of Innsbruck, and Institute for Quantum Optics and Quantum Information, 
Austrian Academy of Sciences, Technikerstrasse 25, 6020 Innsbruck, Austria}

\author{K. Hammerer}
\affiliation{Institute for Theoretical Physics and Institute for Gravitational Physics, Leibniz University Hannover, Callinstrasse 38, 30167 Hannover, Germany}

\author{M.~T. Rakher}
\author{M. Korppi}
\author{A. J\"ockel}
\author{P. Treutlein}
\affiliation{Department of Physics, University of Basel, Klingelbergstrasse 82, 4056 Basel, Switzerland}

\begin{abstract}
We discuss a hybrid quantum system where a dielectric membrane situated inside an optical cavity is coupled to a distant atomic ensemble trapped in an optical lattice. 
The coupling is mediated by the exchange of sideband photons of the lattice laser, and is enhanced by the cavity finesse as well as the square root of the number of atoms. 
In addition to observing coherent dynamics between the two systems, one can also switch on a tailored dissipation by laser cooling the atoms, thereby allowing for sympathetic cooling of the membrane. The resulting cooling scheme does not require resolved sideband conditions for the cavity, which relaxes a constraint present in standard optomechanical cavity cooling. 
We present a quantum mechanical treatment of this modular open system which takes into account the dominant imperfections, and identify optimal operation points for both coherent dynamics and sympathetic cooling. In particular, we find that ground state cooling of a cryogenically pre-cooled membrane is possible for realistic parameters.
\end{abstract}

\pacs{37.30.+i, 07.10.Cm}

\maketitle

\section{Introduction}

Micromechanical resonators coupled to cold atoms have recently attracted much interest as a
promising class of hybrid quantum systems \cite{Wallquist2009,Hunger2011, Treutlein2012}. On the one hand, advances in nanoscale fabrication techniques
have led to the 
development of high-quality micromechanical resonators in various forms, driven in particular by the interest in preparing quantum states of macroscopic devices \cite{Marshall2003}. An attractive feature of these resonators is that they may be functionalized by attaching additional elements, which enables optical, magnetic, or charge-based coupling to other systems \cite{Treutlein2012}.
Many such devices are optimized for the interaction with light in cavity-optomechanical setups \cite{Kippenberg2008,Marquardt2009,Favero2009,Aspelmeyer2010}, which, e.g., enable optomechanical laser cooling
\cite{WilsonRae2007, Marquardt2007, Metzger2004,Gigan2006,Arcizet2006,Kleckner2006,Corbitt2007,Thompson2008,Schliesser2008,Wilson2009}
of the mechanical resonator to its ground state \cite{OConnell2010,Teufel2011,Chan2011}. 
In addition to being useful for applications in metrology 
\cite{Milburn2011}, 
the control of these devices on a single quantum level will also enable 
addressing questions
in quantum foundations \cite{Marshall2003, RomeroIsart2011, Pikovski2012} 
and applications related to quantum information \cite{Zhang2003, Ukram2010, Rabl2010, Stannigel2010,Taylor2011, Chang2011, Wang2012, Tian2012, Agarwal2012,Stannigel2012}.
%
%
%
\begin{figure}[t]
\centering
\includegraphics[width=0.48\textwidth]{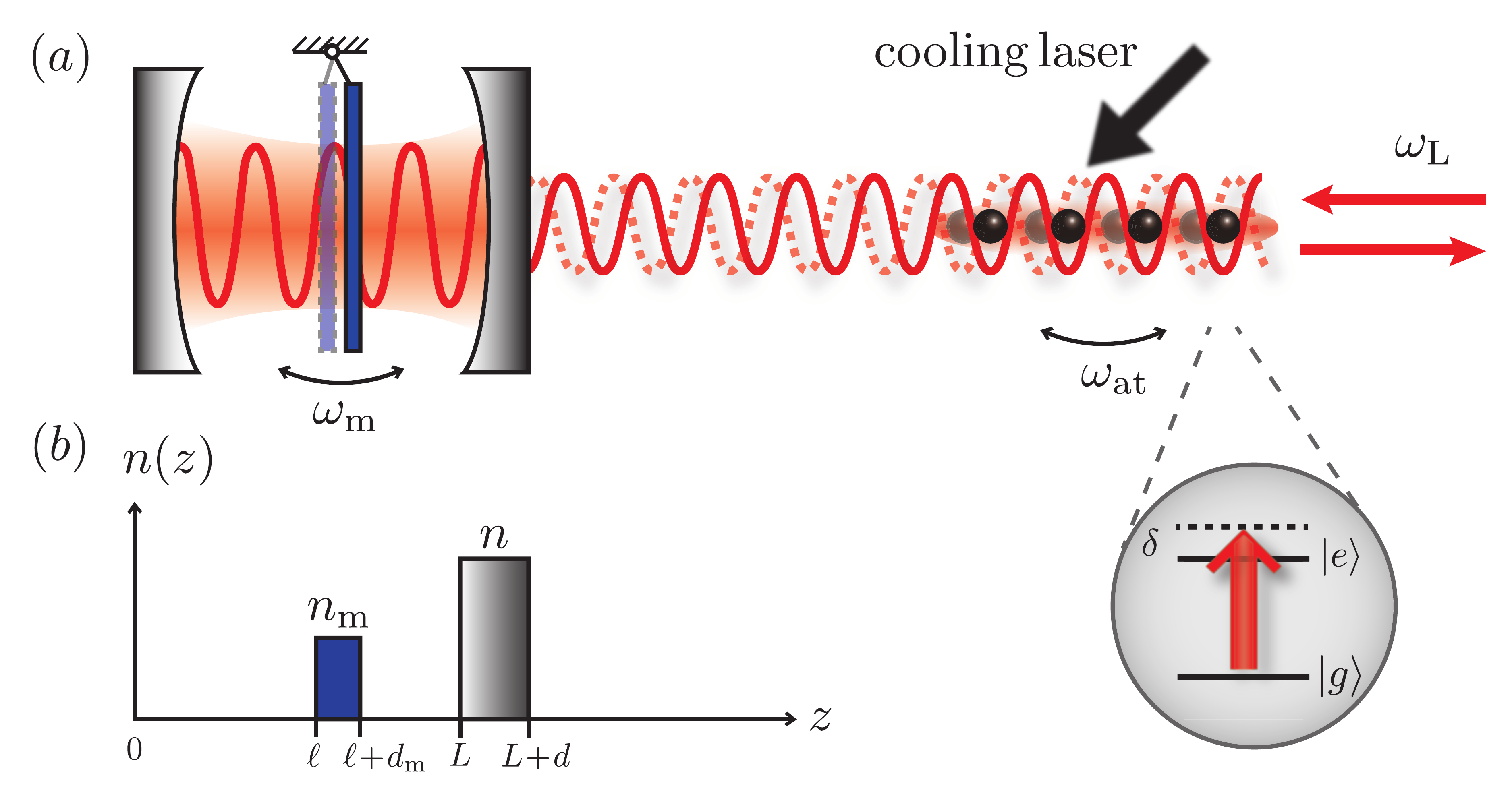}
\caption{(a) Micromechanical membrane in a cavity coupled to a distant atomic ensemble by means of a laser (red, frequency $\omega_\rmL$). The motion of the membrane shifts the optical lattice dipole trap of the atoms, while the motion of the atoms changes the radiation pressure on the membrane.
Inset: The atoms are modeled as two-level systems, which are far off-resonant with respect to the laser.
(b) Refractive index profile used in the one-dimensional model (\eeqref{eq:n_of_z}). }
\label{fig:setup}
\end{figure}
On the other hand, cold atomic gases are systems over which an unprecedented level of quantum control has been reached, involving both internal and external (motional) degrees of freedom \cite{Chu2002}.
Here, a comprehensive toolbox for cooling, state preparation and readout is available, profiting from ongoing efforts in metrology \cite{Cronin2009,Derevianko2011} and quantum simulation \cite{Bloch2012}. These efforts have also enabled an atomic realization of optomechanics \cite{Kanamoto2010, Purdy2010, Brennecke2008}.

In light of these developments, a combination of mechanical and atomic systems promises new ways to control and manipulate the mechanical resonator via the atoms, but may also lead to novel ways of realizing interesting atomic dynamics. 
Along these lines, direct \cite{Treutlein2012,Hunger2011} and optical \cite{Meiser2006,Genes2008,Ian2008,Hammerer2009b,Hammerer2009a,Paternostro2010,Hammerer2010} coupling schemes between atoms and micromechanical systems have been proposed, and first experiments have been realized \cite{Wang2006,Hunger2010,Camerer2011}.
Many of these setups pose great experimental challenges as they require the combination of a functionalized mechanical resonator with atomic trapping and manipulation techniques within the same vacuum chamber or cryostat and, possibly, within the same optical cavity.

A modular setup that circumvents the need for integration of atoms and mechanics within the same vacuum chamber has recently been proposed \cite{Hammerer2010} and later realized \cite{Camerer2011} by some of the authors. The experiment demonstrates a light-mediated coherent coupling of the vibrations of a micromechanical membrane to the center-of-mass motion of an atomic ensemble, which is kept at a distance ($1$~m in the experiment \cite{Camerer2011}) in a separate vacuum chamber.
Motivated by these experimental developments, we analyze in this work 
a more sophisticated setup as displayed in \figref{fig:setup}(a).
Here, the membrane is 
enclosed in an optical cavity, as in the standard `membrane-in-the-middle' configuration \cite{Thompson2008}, while the atoms are trapped 
\emph{outside} of the cavity in the optical lattice potential provided by the light field reflected off the cavity.
Adding the cavity 
preserves the modularity, while the coherent coupling between the two systems is enhanced by the cavity finesse in addition to a collective enhancement by the square root of the number of atoms.
As we will show, the coupling strength can be leveraged so as to exceed the various decoherence rates. First of all, this coherent coupling enables a resonant exchange of excitations between the two systems.
Furthermore, one can switch on laser cooling of the atomic motion \cite{Kerman2000}, thereby effectively tailoring the dissipation of the membrane. This leads to a sympathetic cooling scheme for the membrane which is analogous to standard optomechanical cavity cooling \cite{WilsonRae2007,Marquardt2007}, with the role of the cavity played by the atoms. However, 
in addition to being tunable, our scheme does not require the cavity linewidth to be much less than the mechanical frequency and thus relaxes the experimental requirements on the cavity.

The aim of this work is to provide a rigorous, quantum mechanical treatment of the proposed setup.
In addition to deriving the exact form of the coherent coupling, 
this also yields a consistent description of quantum noise in the problem such that we are able to describe the dynamics of the system on the level of a few quanta.
In particular, we analyze the possibility of observing quantum coherent excitation exchange between the two systems, and address the question whether the membrane can be cooled to its quantum mechanical ground state by laser cooling the atoms. Our results allow us to identify optimal experimental parameters for both scenarios.

The remainder of this paper is structured as follows: In \secref{sec:quantummodel}, we first give a simple semi-classical estimate for the coupling, and then present a full quantum model of the problem from which we obtain a linearized interaction of atoms and membrane with the light field. Section \ref{sec:effDyn} then deals with eliminating the light field to obtain an effective description of atoms and membrane alone, which includes the dominant sources of decoherence. The resulting effective dynamics is subsequently discussed in \secref{sec:discussion}, where we analyze the scenarios of coherent dynamics (\secref{sec:cohdyn}) and sympathetic cooling of the membrane (\secref{sec:sympcool}). Finally, \secref{sec:exprel} comments on the experimental realization of our proposal, and concluding remarks can be found in \secref{sec:conclusions}.

\section{Model}
\label{sec:quantummodel}

We consider a system as shown in \figref{fig:setup}(a), where a mechanical resonator  residing in a single-sided cavity of finesse $\mF$ is coupled to the motion of a distant atomic ensemble \emph{outside} the cavity. 
Before discussing our full quantum model below, we present a brief estimate of the expected coupling based on a quasi-static picture:
the membrane is taken to be a standard SiN membrane with a resonance at frequency $\omega_\rmm$ and effective mass $M$, such that its zero point motion is given by $l_\rmm=\sqrt{\hbar/M\omega_\rmm}$. Its coupling to the atoms is mediated by a laser beam of wavelength $\lambda_\rmL$, which is reflected off the cavity and  forms an optical lattice for the atoms by means of the dipole force \cite{Grimm2000}.  We assume the lattice to be deep enough to treat its wells as independent harmonic traps of frequency $\omega_{\rm at}\approx\omega_\rmm$. If the laser is resonant with the cavity, then a displacement of the membrane by an amplitude $l_\rmm$ corresponding to its 
zero point motion leads to a phase-shift $\delta\phi\sim \mF l_\rmm/\lambda_\rmL$ of the reflected light.  For a single atom, the resulting shift of the standing wave pattern outside the cavity would lead to a  force $\propto m\omega_{\rm at}^2 \delta\phi/k_\rmL$, where $m$ is the mass of one atom and $k_\rmL$ the wavenumber of the laser. However, since all atoms are affected equally the membrane couples to the center-of-mass motion of the atoms, which, on a single quantum level, leads to a collectively enhanced coupling rate of the form \cite{Hammerer2010,Camerer2011}
\begin{align}
\label{eq:estimate}
g\propto m \omega_{\rm at}^2 \frac{\delta\phi}{k_L} \times \frac{l_\rmat \sqrt{N}}{\hbar} \propto \omega_{\rm at} \sqrt{\frac{m}{M}}\, \mF \, \sqrt{N}\,.
\end{align}
Here, $N$ is the number of atoms and $l_{\rmat}=\sqrt{\hbar/m\omega_\rmat}$ is the atomic zero point motion. This estimate shows that the small mass ratio $m/M$ can be compensated by both a large atom number and a high finesse.
In reverse, if the atoms depart from their equilibrium positions, then their restoring force originates from redistribution of photons between the right- and left-propagating components of the standing wave. As a result, the intensity inside the cavity changes and thus the radiation pressure on the membrane such  that it experiences a force. The quantum model presented in the following will confirm the above estimate.

\subsection{Quantum Description}
\label{sec:Hamiltonian}

We describe our setup by a one-dimensional (1D) model with Hamiltonian
\begin{align}
\label{eq:Htot}
H&=H_{\textnormal{m}}+H_{\textnormal{at}}+H_{\textnormal{field}}+H_{\textnormal{m-f}}+H_{\textnormal{at-f}}+H_{\textnormal{dec}}\,.
\end{align}
Here, $H_\rmm$, $H_\rmat$ and $H_{\textnormal{field}}$ describe the free evolution of the mechanics, the external atomic degrees of freedom, and the field modes, respectively. Further, $H_{\textnormal{m-f}}$ represents the coupling between the mechanical oscillator and the field modes, and $H_{\textnormal{at-f}}$ the interaction between atoms and field modes. The last term in \eeqref{eq:Htot} summarizes decoherence effects, which will be introduced explicitly in \secref{sec:decoherence}.

The first term in \eeqref{eq:Htot} describes the harmonic vibrations of the mechanical resonator and reads
\begin{align}
H_{\textnormal{m}}
= \frac{p_\rmm^2}{2 M}+\frac{M \om_\rmm^2 z_\rmm^2}{2}
\equiv \hbar \om_\rmm \left( a_\rmm^\dag a_\rmm +\frac{1}{2}\right),
\label{eq:Hm}
\end{align}
where $z_\rmm$, $p_\rmm$ and $a_\rmm$ are the position, momentum and annihilation operators, respectively. The second term in \eeqref{eq:Htot} is given by $H_{\textnormal{at}}=\sum_{j=1}^N p_j^2/ 2m$ and represents the kinetic energy of the atoms, where the $p_j$ are the momenta of the individual atoms.
The free field Hamiltonian reads $H_{\textnormal{field}}=\int\rmd\om\, \hbar\om\,b_\om^\dag b_\om$, where the mode operators obey the commutation relation $[b_\om,b_{\om'}^\dag]=\delta(\om-\om')$, and all frequency integrals are understood to run over a bandwidth $2\theta$ around the laser frequency $\om_L$ introduced below. In the spirit of a `modes of the universe' approach we take care of the cavity by explicitly including it in the mode functions $u_\omega(z)$, which are associated with the $b_\omega$ and defined by writing the positive frequency parts of the fields as \cite{QuantumNoise}
\begin{align} \label{eq:radfield}
\hat E^{(+)}(z)&=i\int\rmd\om\,\mE_\om u_\om(z) b_\om\,,  \\
\hat B^{(+)}(z)&=\frac{i}{c}\int\rmd\om\,\mE_\om \bar u_\om(z) b_\om\, .
\end{align}
Here, $\mE_\om=\sqrt{\hbar\om/\pi\epsilon_0 c \mA}$, $\bar u_\om(z)=\partial_z u_\om(z)/ik$ and $\mA$ is the cross sectional area of the modes.
In the following we discuss the particular case of a `membrane-in-the-middle' setup \cite{Thompson2008} based on a single-sided cavity, as shown in \figref{fig:setup}(a). However, other arrangements can be treated analogously by using the appropriate mode functions. This is shown explicitly in \appref{app:mirror} for a setup with a movable end mirror instead of the membrane.

We model the membrane and the semi-transparent end mirror of the cavity by dielectric slabs at positions $\ell$ and $L$, respectively \cite{Thompson2008, Jayich2008}. The wave equation for the mode functions thus has to be solved for the refractive index profile
\begin{equation}
\label{eq:n_of_z}
n(z) =\begin{cases}
n_\rmm& \text{if } z \in [\ell, \ell+d_\rmm] \\
n & \text{if } z\in[L, L+d]\\
1 & \textrm{otherwise}
\end{cases},
\end{equation}
which is illustrated in \figref{fig:setup}(b), and with the boundary condition of a vanishing field at $z=0$. The latter corresponds to the limiting case of an ideal
mirror at position $z=0$. In the above equation, thickness ($d,d_\rmm$) and refractive index ($n,n_\rmm$) of each slab determine the associated amplitude reflection and transmission coefficients, which read \cite{Born1999}
\begin{align}
\mathfrak{r}&=\frac{(n^2-1) \sin(k d n)}{2 i n \cos(k d n)+(1+n^2) \sin(k d n)}\,, \label{eq:reflmirror}\\
\mathfrak{t}&=\frac{2 i n}{2 i n \cos(k d n)+(1+n^2) \sin(k d n)}\,, \label{eq:transmirror}
\end{align}
for the end mirror and analogous expressions hold for the coefficients $\mathfrak{r}_\rmm$ and $\mathfrak{t}_\rmm$ of the membrane.
As a result, the properly normalized mode functions read
\begin{align}
\label{eq:membraneMF}
u_\om(z)&=\frac{1}{2i} \begin{cases}
T_\om( e^{i k z}- e^{-i k z}) & \text{if }0<z<\ell \\
T_\om\left(\frac{e^{i k z}}{A_\om^*} -\frac{e^{-i k z}}{A_\om}\right) & \text{if }\ell+d_\rmm<z<L \\
\frac{T_\om}{T_\om^*} e^{i k z}-e^{-i k z} & \text{if }z>L+d \\
\end{cases},
\end{align}
where we have defined the coefficients
\begin{align}
A_\om&=\frac{\mft_\rmm e^{-i k d_\rmm}}{1-\mfr_\rmm e^{2ik\ell}}\equiv|A_\om|e^{i\varphi_\omega^\prime}\,,\\
T_\om&=A_\om \frac{\mft e^{-ikd}}{1-\mfr e^{2 i (k \rmL+\varphi'_\om)}}
\equiv |T_\om|e^{i\varphi_\omega}\label{eq:amplA}.
\end{align}
Clearly, $\varphi_\omega$ is the phase by which the standing wave pattern outside the cavity is shifted.
In the limit of an almost transparent membrane and a high reflectivity end mirror, given by $\mfr_\rmm \rightarrow 0$, $|\mfr|\rightarrow 1$, the factor $T_\om$ represents the response of an empty single-sided Fabry-P\'erot cavity with well-defined resonance frequencies $\om_\nu=\nu \pi c/L$, where $\nu=1,2,\ldots$. Around the latter, $\sabs{T_\om}^2$ can be approximated by a Lorentzian, i.e.
\begin{align}
\sabs{T_\om}^2\Big\vert_{\om\approx\om_\nu}
\approx & \frac{2\mF}{\pi}\frac{\kappa^2}{\kappa^2+(\om-\om_\nu)^2}.
\label{eq:lorentzian}
\end{align}
Here, $\kappa=c(1-|\mfr|) / 2L |\mfr|^{1/2}$ is half of the cavity linewidth and $\mF=\pi c /2 \kappa L$ is the finesse, which we assume to be limited by the semi-transparent end mirror.  
For small displacements $z_\rmm$ the radiation pressure interaction between membrane and field in \eeqref{eq:Htot} is given by \cite{Hammerer2010,Yamamoto2010}
\begin{align}
H_{\textnormal{m-f}}=\frac{\epsilon_0 (n_\rmm^2-1) \mA}{2} &\left[ \hat E^{(-)}(\ell+d_\rmm) \hat E^{(+)}(\ell+d_\rmm)\right. \notag \\
&\left. -\hat E^{(-)}(\ell)\hat E^{(+)}(\ell)\right]  z_\rmm \,,
\label{eq:Hintmf_membrane}
\end{align}
which we will evaluate below.

Finally, $H_{\textnormal{at-f}}$ in \eeqref{eq:Htot} describes the dipole interaction of the atoms with the light field. We model the atoms as two-level systems of transition frequency $\omega_{\rm eg}$, from which the frequency $\om_\rmL$ of the trapping laser is far detuned by $\delta=\om_\rmL-\om_{\rm eg}$ (see inset in \figref{fig:setup}(a)). We further anticipate that all relevant field modes will lie within a bandwidth $2\theta\ll|\delta|$ around the laser frequency, such that we can eliminate the upper level of the atoms and describe the atom-field interaction by an AC-Stark shift for atoms in the ground state
\begin{align}
H_{\textnormal{at-f}}= \frac{\mu^2}{\hbar\delta} \sum_{j=1}^N \hat E^{(-)}(z_j) \hat E^{(+)}(z_j),
\label{eq:Hintatf}
\end{align}
where $\mu$ is the atomic dipole moment and the $z_j$ are the positions of the atoms.

\subsection{Linearization Around Laser Drive}
\label{subsec:LinearLaser}

We now introduce the laser displayed in \figref{fig:setup}(a), which provides the optical lattice for the atoms and  mediates the coupling between the atoms and the mechanical resonator. To do so, we carry out the replacement for the field mode operators
\begin{align}
\label{eq:displacement}
b_\om \rightarrow b_\om + \delta(\om-\om_L) \,  \alpha \, e^{-i\om_L t},
\end{align}
where the amplitude $\alpha$ is related to the running wave power $P=\hbar \om_\rmL \alpha^2/2\pi$ of the incident laser. In the following we assume $\alpha \gg 1$, which allows us to linearize the interaction Hamiltonians $H_{\rm m-f}$ and $H_{\rm at-f}$ around the laser by only keeping terms of order $\alpha$ and $\alpha^2$.

\subsubsection{Membrane-Field Interaction}
\label{sec:membranefieldint}

We start with the membrane-field interaction $H_{\textnormal{m-f}}$ by inserting the field expansion in \eeqref{eq:radfield} together with the mode functions given in \eeqref{eq:membraneMF} into \eeqref{eq:Hintmf_membrane},  and then apply the above replacement to the field operators.
The contribution $\propto\alpha^2$ yields a constant force on the membrane that can be accounted for by a redefinition of its equilibrium position.
The contribution linear in $\alpha$ provides the relevant interaction between mechanics and light field.
For convenience we introduce
$\lambda_{l,\om}=u_\om(\ell)$ and $\lambda_{r,\om}=u_\om(\ell+d_\rmm)$
for the mode functions such that in a rotating frame with respect to the laser frequency the linearized membrane-field interaction $H_{\textnormal{m-f}}$ is given by the contribution linear in $\alpha$:
\begin{align}
H_{\textnormal{m-f}}&=\frac{ \epsilon_0 (n_\rmm^2-1)\mA \alpha }{2} \!\int \! {\rm d}\om \, \Big[ \{\lambda^*_{r,\om_\rmL} \lambda_{r,\om}-\lambda^*_{l,\om_\rmL} \lambda_{l,\om}\} b_\om  \notag \\
& \qquad\qquad -\{ \lambda^*_{r,\om} \lambda_{r,\om_\rmL}-\lambda^*_{l,\om}\lambda_{l,\om_\rmL} \} b_\om^\dag\Big] \mE_\om \mE_{\om_\rmL}  z_\rmm\,.
\end{align}
At this point we need to make assumptions on the frequency-dependence of the mode functions $u_\om(z)$ in order to proceed. We focus on the case of small membrane reflection $|\mfr_\rmm|$, such that the cavity response is not altered significantly by the presence of the membrane and assume that the laser is resonant with a cavity mode. As motivated in \secref{sec:effDyn} below, we further assume a `bad cavity' regime, where the cavity half-linewidth $\kappa$ is much larger than the separation of the relevant sidebands from the laser frequency $\om_\rmL$ (see \figref{fig:badcavity}). In this case we may approximate $|T_\om|\approx|T_{\om_\rmL}|$, such that  $u_\om(z) \approx u_{\om_\rmL}(z)$, and by using $\mE_\om \approx \mE_{\om_\rmL}$ we obtain
\begin{align}
H_{\textnormal{m-f}}&\approx\hbar g_\rmm  \int\!  \frac{\rmd\om}{\sqrt{2\pi}} (b_\om e^{i \Delta \varphi_\om} + b_\om^\dag e^{-i \Delta \varphi_\om} )(a_\rmm+a_\rmm^\dag)\,,
\label{eq:Hmf_lin}
\end{align}
with the phase difference $\Delta \varphi_\om=\varphi_\om-\varphi_{\om_\rmL}$ and the membrane-field coupling constant
\begin{align}
g_\rmm=\frac{\sqrt{\pi}}{2\hbar} \epsilon_0 (n_\rmm^2-1)\mA\alpha \mE_{\om_\rmL}^2 l_\rmm (|\lambda_{r,\om_\rmL}|^2-|\lambda_{l,\om_\rmL}|^2)\,.
\end{align}
In the limit $k_\rmL d_\rmm\rightarrow 0$ of a thin membrane and small but constant reflection $|\mfr_\rmm|\approx k_\rmL d_\rmm (n_\rmm^2-1)/2$ we can write the membrane-field coupling constant as
\begin{align}
g_\rmm&=\frac{\alpha k_\rmL l_\rmm}{ \sqrt{ \pi}} \, \sabs{\mfr_\rmm} \sin(2 k_\rmL \ell)\,\sabs{T_{\om_\rmL}}^2 \label{eq:gm1} \\
&= \frac{\alpha k_\rmL l_\rmm}{ \sqrt{ \pi}}\, \sabs{\mfr_\rmm}  \,  \frac{2 \mF}{\pi} \,, 
\label{eq:gm2}
\end{align}
where we assumed $\sin(2 k_\rmL \ell)=1$ in the second line. This corresponds to placing the membrane on the slope of the intra-cavity intensity, where the coupling is maximal. In contrast, at points of zero intensity the coupling vanishes and higher orders in $z_\rmm$ would have to be considered \cite{Thompson2008}. However, these points will not be relevant in this work. Note that the coupling in \eeqref{eq:gm2} is enhanced by the finesse and that a similar calculation for a cavity with movable end mirror yields exactly the same interaction in \eeqref{eq:Hmf_lin}, but with 
$g_\rmm \rightarrow g^\prime_\rmm= 2\alpha k_\rmL l_\rmm   \mF / \pi^{3/2}$, as detailed in \appref{app:mirror}.

\subsubsection{Atoms-Field Interaction}
\label{sec:atfinteraction}

The next step is the expansion of the AC-Stark shift Hamiltonian $H_{\textnormal{at-f}}$ in powers of $\alpha$. We insert \eeqref{eq:radfield} with the mode functions of \eeqref{eq:membraneMF} into \eeqref{eq:Hintatf} and apply the replacement \eeqref{eq:displacement}. The resulting contribution proportional to $\alpha^2$ provides an optical lattice trap that is seen by each atom and that is described by the potential $V(z_j)=V_0 \sin^2(k_\rmL z_j+\varphi_{\om_\rmL})$ with lattice depth $V_0=\mu^2\mE_{\om_\rmL}^2\sabs{\alpha}^2/\hbar\delta$. The minima $\bar z_j$ for blue detuning ($\delta > 0$) are determined by the condition $k_\rmL \bar{z}_j+\varphi_{\om_\rmL}=\nu \pi$, where $\nu=1,2,\ldots$.
Combined with the kinetic energy of the atoms in $H_{\textnormal{at}}$ a harmonic approximation around the trap minima defines the atomic trap frequency $\om_\rmat$  via $m \om_\rmat^2/2=V_0 k_\rmL^2$.
The contribution linear in $\alpha$ yields the interaction between light field and atomic ensemble.
By applying a Lamb-Dicke expansion around the equilibrium positions of the atoms $\bar{z}_j$, we obtain a linear interaction between the field modes $b_\om$ and the fluctuations around the equilibrium position of each atom given by $z_j=l_\rmat(a_j^\dag+a_j)/\sqrt{2}$, where $[a_j,a_j^\dag]=1$. 
In carrying out the remaining sum in \eeqref{eq:Hintatf} we replace the $\bar z_j$ by the center-of-mass position $\bar z$ of the whole ensemble, which is a good approximation if $\om_\rmat L_\rmat/c \ll 1$, where $L_\rmat$ is the linear extension of the ensemble. For typical experimental parameters this assumption is fulfilled in general, see \secref{sec:exprel}. As a result, the light field only couples to the atomic center-of-mass mode $a_\rmat=\sum_j a_j/\sqrt{N}$ according to
\begin{align}
H_{\textnormal{at-f}}&= \hbar g_\rmat \int \! \frac{\rmd \om}{\sqrt{2\pi}} \,  \left( b_\om e^{i \Delta \varphi_\om}  + b_\om^\dag e^{-i \Delta \varphi_\om} \right)(a_\rmat+a_\rmat^\dag) \notag \\
&\qquad \qquad \qquad \times \sin{\left[(k-k_L)\bar{z} + \Delta\varphi_\omega \right]},
\label{eq:Hatf_lin}
\end{align}
where the atom-field coupling
\begin{align}
g_\rmat= \frac{\om_\rmat}{2 \alpha k_L l_\rmat} \sqrt{\pi}\sqrt{N}
\label{eq:gat}
\end{align}
is enhanced by the square-root of the number of atoms.

\subsubsection{Final Linearized Hamiltonian}

In summary, the final linearized Hamiltonian contains interactions between the field modes and the mechanical and atomic center-of-mass motion. We neglect the constant contribution in \eeqref{eq:Hm} and together with \eeqref{eq:Hmf_lin} and \eeqref{eq:Hatf_lin} the linearized Hamiltonian in a frame rotating at the laser frequency reads
\begin{align}
\label{eq:Hlin}
H&_\textrm{lin}= \hbar \om_\rmm a_\rmm^\dag a_\rmm +  \hbar \om_\rmat a_\rmat^\dag a_\rmat
+\int\rmd\om\, \hbar\Delta_\om\,b_\om^\dag b_\om \\
&+\hbar g_\rmm \! \int \! \frac{\rmd\om}{\sqrt{2\pi}} (b_\om + b_\om^\dag )(a_\rmm+a_\rmm^\dag)\nonumber\\
&+\hbar g_\rmat\! \int  \! \frac{\rmd\om}{\sqrt{2\pi}}  ( b_\om  + b_\om^\dag )(a_\rmat+a_\rmat^\dag)\sin(\Delta_\om\bar{z}/c+\Delta \varphi_\om) .\nonumber
\end{align}
Here, $\Delta_\om=\om-\om_\rmL$ and we have absorbed the phase factors $ e^{\pm i \Delta \varphi_\om}$ appearing in Eqs.\,\eqref{eq:Hmf_lin} and \eqref{eq:Hatf_lin} into the field operators $b_\omega$. The effect of the cavity response $T_\om$ enters both via the membrane-field coupling $g_\rmm$, which depends on $\abs{T_{\om_\rmL}}^2$, and the phase $\varphi_\omega={\rm arg} \{T_\omega\}$ contained in $\Delta\varphi_\omega$.

\section{Effective Dynamics}
\label{sec:effDyn}

\begin{figure}[t]
\centering
\includegraphics[width=0.4\textwidth]{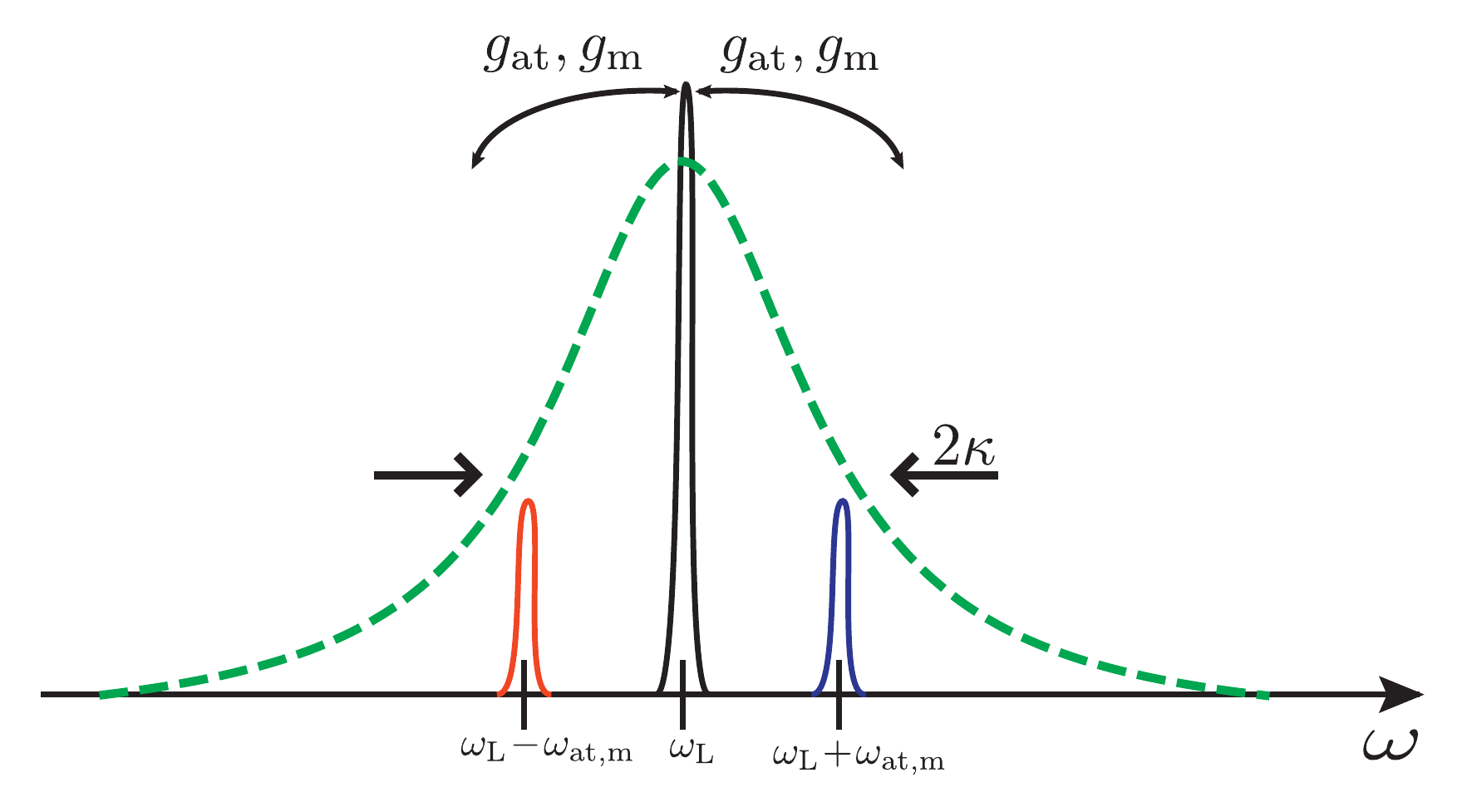}
\caption{Light fields involved in mediating the effective coupling. The laser-enhanced interactions (coupling constants $g_\rmm$ and $g_\rmat$) correspond to a conversion between laser photons $\om_\rmL$ (black) and sideband photons $\om_\rmL \pm \om_{\rmat, \rmm}$ (blue/dark and red/light). The response profile of the cavity is indicated by the green dashed line and easily accommodates both sidebands in the `bad cavity' regime discussed in the text.}
\label{fig:badcavity}
\end{figure}

We are interested in deriving the effective equations of motion (EOMs) describing the coupling between atoms and mechanical oscillator, which are obtained by eliminating the light field in a Born-Markov approximation. Motivated by the experimental considerations presented below in \secref{sec:exprel}, we focus in the following on the `bad cavity' regime, where the cavity half-linewidth $\kappa$ is much larger than the atomic and 
mechanical frequency, i.e. $\kappa \gg \omega_{\rm at},\omega_{\rm m}$ as depicted in \figref{fig:badcavity}. Also assuming the weak-coupling conditions $\omega_{\rm at},\omega_{\rm m} \gg g_{\rm at}^2,g_{\rm m}^2,g_{\rm at}g_{\rm m}$ we find that the resonant terms in \eeqref{eq:Hlin} involve field operators at the sideband frequencies $\om=\om_\rmL\pm \om_{\rmat,\rmm}$, and we choose $\om_\rmat\approx \om_\rmm$ to ensure resonant interactions. We further assume that the laser is resonant with a cavity mode, i.e. $\om_\rmL \approx \om_\nu$ as shown  in \figref{fig:badcavity}, which yields a high intra-cavity intensity, but also ensures that both sidebands enter the cavity equally well, i.e., that there is no spectral filtering.
In the described regime the cavity adiabatically follows the mechanical oscillator and the atoms and needs not to be treated as an independent degree of freedom. 
The opposite `resolved sideband' limit, where $\kappa \ll \om_\rmm,\om_\rmat$, would require a different treatment and is not considered in this work.
Finally note that the assumptions described here already entered in the derivation of \eeqref{eq:Hlin}.

\subsection{Elimination of the Light Field}
\label{sec:elimination}
We start by stating the complete hierarchy of timescales for our calculation
\begin{align}
\label{eq:HierarchyTimescales}
\delta \gg \theta\gg\frac{1}{\tau}, \kappa \gg \om_\rmat,\om_\rmm \gg g_\rmm^2, g_\rmat^2, g_\rmm g_\rmat,
\end{align}
where $\tau=\bar{z}/c$ is the propagation time between mechanics and atomic ensemble. The left-most inequality is needed to model the atom-light interaction by an AC-Stark shift as described in \secref{sec:Hamiltonian}.
We assume a broad spectrum of field modes such that its bandwidth $2\theta$ is large compared to all frequencies associated with the system evolution. Below, this allows us to distinguish the temporal order of the light field's interactions with the atoms and the mechanical resonator.
Motivated by experimental parameters (see \secref{sec:exprel}), we assume that the retardation time between the two systems is short compared to the system's evolution, i.e. $1/\tau \gg \om_\rmat,\om_\rmm$. Finally, we have the bad cavity assumption as well as the weak-coupling assumptions $\omega_{\rm at},\omega_{\rm m} \gg g_{\rm at}^2,g_{\rm m}^2,g_{\rm at}g_{\rm m}$ as discussed in the the previous paragraph.

To derive the effective EOMs for mechanical oscillator and atoms we start with the Heisenberg EOMs for the quadrature operators obtained from the linearized Hamiltonian given in \eeqref{eq:Hlin} (see \appref{app:elimination} for details). We then formally integrate the EOMs for the field quadratures. By inserting the resulting expressions into the Heisenberg equations for mechanics and atoms and applying the Born-Markov approximation, we obtain effective quantum Langevin equations (QLEs) with interactions retarded by a time $\tilde{\tau}=\tau+1/\kappa$. The additional contribution $1/\kappa$ represents the retardation due to the cavity and follows from the Taylor expansion of $\Delta \varphi_\om$ around the cavity resonance. From \eeqref{eq:HierarchyTimescales} we see that $\omega_\rmat \tilde{\tau}\ll 1$, so that we can neglect these retardations in the following, and thus obtain effective QLEs that are local in time:
\begin{subequations}
\label{eq:QLEs}
\begin{align}
\dot p_\rmm &= -\om_\rmm x_\rmm +g x_\rmat + \sqrt{\gamma_\rmm^{\rm diff}}\,F_\rmm(t) \label{eq:QLE1}
\\
\dot x_\rmm &= \om_\rmm p_\rmm\\
\dot p_\rmat &= -\om_\rmat x_\rmat + g  x_\rmm  \label{eq:QLE3}
\\
\dot x_\rmat &= \om_\rmat p_\rmat \,.
\end{align} \end{subequations}
Here, $F_\rmm(t)$ and $\gamma_\rmm^{\rm diff}$ are defined below, and $g=2 g_\rmat g_\rmm$ is the effective coupling between the two systems. From the above equations we can read off the corresponding effective Hamiltonian
\begin{align}
\label{eq:Heff}
H_{\rm eff}=\hbar \om_\rmm a_\rmm^\dag a_\rmm +  \hbar \om_\rmat a_\rmat^\dag a_\rmat
-\hbar g x_\rmm x_\rmat\,,
\end{align}
where the effective coherent coupling $g$ contains the desired enhancements and is explicitly given by 
\begin{align}
g=2g_\rmat g_\rmm = &\om_\rmat \frac{l_\rmm}{l_\rmat} \sabs{\mfr_\rmm}   \sqrt{N} \,\frac{2 \mF}{{\pi}}.
\label{eq:effectivecoupl}
\end{align}

The noise operator $F_\rmm(t)$ appearing in \eeqref{eq:QLE1} is Hermitian and represents the radiation pressure noise originating from the intra-cavity light field.
It is characterized by the two-time correlation function
\begin{align}
\smean{F_\rmm(t)F_\rmm(t')}  &= \delta_\theta(t-t'),
\end{align}
and the associated diffusion rate
\begin{align}
\label{eq:gammamdiff}
\gamma_\rmm^\textnormal{diff}=2 g_\rmm^2=\frac{4 P}{M c^2}\frac{\om_\rmL}{\om_\rmm} \sabs{\mfr_\rmm}^2 \left(\frac{2 \mF}{\pi}\right)^2
\end{align} 
appearing in \eeqref{eq:QLE1}. The expression $\delta_\theta(t-t')$ denotes a representation of the $\delta$-function that is peaked on the timescale $1/\theta$.
As explained in \appref{app:elimination} this momentum diffusion rate appears as a fundamental limitation in any optomechanical system, and it was recently observed in an experiment \cite{Purdy2012}.

\subsection{Decoherence and Imperfections}
\label{sec:decoherence}
In the following we will discuss the main decoherence channels of the mechanical oscillator and the atoms. 
In addition to the radiation pressure noise derived above, these include thermal decoherence due to finite support temperature and absorption of laser photons in the membrane, as well as light-induced momentum diffusion of the atoms.

The mechanical resonator is coupled to a support of temperature $T_0$, which leads to a finite linewidth $\gamma_\rmm$ due to clamping losses, but also to thermal decoherence at rate $\gamma_\rmm N_\rmm\approx \frac{k_B T_0}{\hbar Q_\rmm}$, where $k_B$ is the Boltzmann constant and $Q_\rmm=\om_\rmm/\gamma_\rmm$ the mechanical quality factor.
In addition, the membrane is also heated by absorbing photons from the coupling laser. We model this heating effect by introducing an effective bath temperature $T_{\rm eff}=T_0+\delta T$, where $\delta T= \frac{P_{\rm abs}}{K_{\rm th}}$ is the temperature increase due to absorbed laser power $P_{\rm abs}$ in the center of the membrane, representing a worst-case estimate \cite{Hammerer2009b}. 
The thermal link $K_{\rm th}$ connects the center of the membrane to the support and depends on the thermal conductivity and  the thickness of the membrane, as well as the beam waist of the laser (see \appref{app:abs}). Further, the absorbed power for a membrane sitting on a slope of the intra-cavity intensity is given by $P_{\rm abs} = \mathfrak{a}_\rmm^2 \frac{ 4\mF}{\pi} P$, where $\mathfrak{a}_\rmm^2=1-\mathfrak{r}_\rmm^2-\mathfrak{t}_\rmm^2$ is the power absorption coefficient. The resulting effective bath occupation is thus given by 
\begin{align}
\overline{N}_{\rmm}\approx\frac{k_B T_{\rm eff}}{\hbar \om_\rmm}=\frac{k_B}{\hbar \om_\rmm} \left( T_0+\frac{\mathfrak{a}_\rmm^2}{K_{\rm th}} \frac{4 \mF}{\pi}P\right)\,,
\label{eq:Nmbar}
\end{align}
as  described in more detail in \appref{app:abs}.

Atoms in a 3D standing wave undergo diffusion processes. In our treatment these diffusion processes drop out as an artifact of the 1D model, as already mentioned in \appref{app:elimination} and in Ref.\,\cite{Hammerer2010}. Therefore, we have to add the proper momentum diffusion for atoms in a three-dimensional (3D) standing wave as, e.g., derived in Ref.\,\cite{Gordon1980} and given by $\gamma_\rmat^{\rm diff}= (k_{\rmL} l_\rmat)^2 \gamma_{\rm se} V_0/\hbar \sabs{\delta}$, where $\gamma_{\rm se}$ is the natural linewidth of the transition.

Taking all of these noise sources into account, we can now give the full QLEs describing the effective dynamics of the system:
\begin{align}
\dot{a}_\textrm{m}(t)&=-\frac{i}{\hbar}[a_\rmm, H_{\rm eff}] -\frac{\gamma_\rmm}{2} a_\textrm{m}(t) \notag \\
& \qquad+i\sqrt{\frac{\gamma_\rmm^{\rm diff}}{2}} F_{\textrm{m}}(t) +\sqrt{\gamma_\rmm} \xi(t)\label{eomamgscapp}\\
  \dot{a}_\textrm{at}(t)&=-\frac{i}{\hbar}[a_\rmat, H_{\rm eff}]+i \sqrt{\frac{\gamma_\rmat^{\rm diff}}{2}}F_{\textrm{at}}(t)\,.
\label{eomaatgscapp}
\end{align}
Here, the operator $\xi(t)$ models thermal mechanical noise and is  characterized by $[\xi(t),\xi^\dag(t') ]=\delta(t-t')$ and  $\smean{\xi^\dag(t)\xi(t^\prime)}=\overline{N}_\rmm \delta(t-t^\prime)$ \cite{QuantumNoise}. In analogy to $F_\rmm(t)$, the Hermitian noise operator $F_{\rmat}(t)$ accounts for atomic diffusion and satisfies $\smean{F_{\textrm{at}}(t)F_{\textrm{at}}(t^\prime) }=\delta(t-t^\prime)$.

Finally, we comment on photon loss on the optical path between membrane and atoms. Such loss could be modeled by placing a beam-splitter of suitable reflectivity between the two systems, such that the light fields $b_\om$ are coupled to additional vacuum modes. Integrating out the light fields as above would then yield a reduced coherent coupling, analogous to the situation treated in Refs. \cite{Hammerer2010,Camerer2011}. There, a similar setup without cavity was discussed, and photon loss through the membrane into the $-z$ direction was included (cf. \figref{fig:setup}).

\section{Discussion and applications}
\label{sec:discussion}

In the previous sections, we have reduced the description of our setup to the effective QLEs \eqref{eomamgscapp} and \eqref{eomaatgscapp}, which feature a coherent coupling between atoms and membrane as described by the effective Hamiltonian \eqref{eq:Heff}. However, they also contain a number of thermal and light-induced noise sources, whose impact we discuss in the following for two different scenarios. Our final goal is to establish interesting regimes of operation for the proposed setup.

\subsection{Coherent Dynamics}
\label{sec:cohdyn}

As a first scenario we consider the observation of the coherent dynamics induced by the interaction term in the Hamiltonian \eqref{eq:Heff}. A signature of the coupling is the normal mode splitting $\sim g$, which is visible in the fluctuation spectra of the oscillators, as soon as it exceeds their amplitude decay rates (see Ref. \cite{Groblacher2009} for an analogous discussion in standard optomechanics). 
Note that  normal mode splitting also occurs for classical oscillators and that it is visible also at high temperatures \cite{Groblacher2009}.
This effect could be observed with an additional weak probe laser coupled to the cavity that reads out the membrane vibrations. Analyzing the frequency spectrum of the reflected light would reveal the normal modes. 
In the time domain, the same effect gives rise to an oscillatory exchange of excitations between the membrane and the atomic vibrations. These oscillations could be observed with the probe laser after exciting the membrane with a piezo at frequency $\omega_\rmm$.

\begin{figure}[t]
\centering
\includegraphics[width=0.5\textwidth]{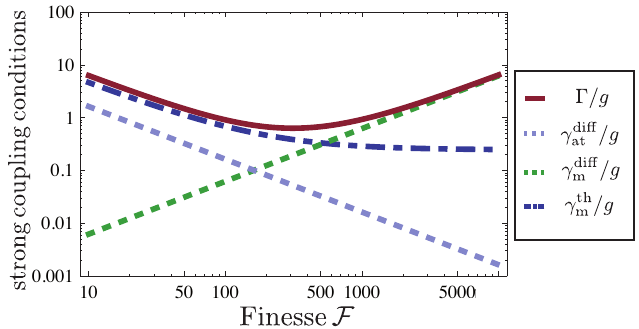}
\caption{Coherent dynamics: We plot the ratios $\gamma_\rmat^{\rm diff}/g$, $\gamma_\rmm^{\rm diff}/g$, and $\gamma_\rmm^{\rm th}/g$  as a function of finesse to illustrate the strong coupling conditions in \eeqref{eq:strongcoupl}. One clearly observes a trade-off between mechanical heating ($\gamma_\rmm^{\rm th}$) and radiation pressure noise ($\gamma_\rmm^{\rm diff}$). Parameters other than $\mF$ are taken from Tab. \ref{tab:parameter}.}
\label{fig:strongcoupling}
\end{figure}

Ultimately, one would like to observe the effects of the coherent coupling on a single quantum level. In this context, the regime $g\ll\om_\rmm\approx\om_\rmat$ appears naturally, where the effective interaction in \eeqref{eq:Heff} can be approximated by a beam-splitter Hamiltonian according to 
\begin{align}
\label{eq:HeffRWA}
H_{\rm eff}\approx\hbar \om_\rmm a_\rmm^\dag a_\rmm +  \hbar \om_\rmat a_\rmat^\dag a_\rmat
-\frac{\hbar g}{2} (a_\rmm^\dag a_\rmat + a_\rmat^\dag a_\rmm )\,,
\end{align}
which enables the \emph{coherent} transfer of single excitations between atoms and membrane. Since all decoherence mechanisms are detrimental in this case, the coupling has to exceed all of the associated rates as expressed by the strong coupling conditions
\begin{align}
g \gg \gamma^{\rm diff}_\rmat, \gamma^{\rm diff}_\rmm,  \gamma_\rmm^{\rm th}\,,
\label{eq:strongcoupl}
\end{align}
where $\gamma_\rmm^{\rm th}=\gamma_\rmm \overline{N}_\rmm$. In order to identify a regime where all of these inequalities can be fulfilled, we plot in \figref{fig:strongcoupling} the ratios $\gamma_\rmat^{\rm diff}/g$, $\gamma_\rmm^{\rm diff}/g$, and $\gamma_\rmm^{\rm th}/g$ as a function of finesse, while fixing all other system parameters to the values of Tab. \ref{tab:parameter}.
Note that $g\propto \mF$, such that the relative importance $\gamma_\rmat^{\rm diff}/g$ of the atomic decoherence decreases as $1/\mF$, although $\gamma_\rmat^{\rm diff}$ is constant. In contrast, the mechanical radiation pressure noise grows quadratically with $\mF$, such that $\gamma_\rmm^{\rm diff}/g \propto \mF$. Finally, $\gamma_\rmm^{\rm th}/g$ scales as $1/\mF$ for small finesses and then saturates due to the laser-induced heating of the membrane for large finesses. The plot shows that the atomic heating is irrelevant for the chosen parameters (which is due to the chosen large detuning $\delta$), whereas there is a clear trade-off between mechanical heating and radiation pressure noise, with the latter dominating at high finesses. To illustrate this, the plot also shows the sum $\Gamma/g$ of the above ratios, where $\Gamma=\gamma_\rmat^{\rm diff}+\gamma_\rmm^{\rm diff}+\gamma_\rmm^{\rm th}$. While $\Gamma/g\lesssim 10$ over the whole plot range, one clearly sees that there is an optimal value of the finesse at $\mF\approx300$, where $\Gamma/g\approx1$. Interestingly, the condition $g\gg \gamma_\rmm^{\rm diff}$ implies an asymmetry in the coupling constants according to $g_\rmat \gg g_\rmm$ (cf. Eqs. \eqref{eq:effectivecoupl} and \eqref{eq:gammamdiff}).

The above estimates show that observing normal mode splitting, as well as quantum coherent dynamics, is possible for present-day parameters. Going further, the interaction in \eeqref{eq:HeffRWA} can be the basis for a quantum state transfer protocol between the two systems (see, e.g., Ref.\,\cite{Wallquist2010}), which could become feasible for a more optimized setup.

\subsection{Sympathetic Cooling}
\label{sec:sympcool}

The versatility of the atoms allows one to switch on dissipation, e.g., by laser cooling their motion \cite{Kerman2000,Camerer2011}. Together with the resonant phonon exchange between mechanics and atoms contained in the Hamiltonian in \eeqref{eq:Heff}, this gives rise to a sympathetic cooling effect for the membrane in analogy to optomechanical cavity cooling.
To describe this effect, we model the laser cooling by introducing an amplitude decay rate $\gamma_\rmat^{\rm cool}/2$ for the atomic center-of-mass mode, such that the QLE \eqref{eomaatgscapp} is replaced by
\begin{align}
\label{eomaatgsc}
  \dot{a}_\textrm{at}(t)&=-\frac{i}{\hbar}[a_\rmat, H_{\rm eff}]-\frac{\gamma_\rmat^{\textnormal{cool}}}{2} a_\textrm{at}(t) \notag \\
  &\qquad +i\sqrt{\frac{\gamma_\rmat^{\rm diff}}{2}}F_{\textrm{at}}(t)+\sqrt{\gamma_{\rmat}^{\textnormal{cool}}} F_\textnormal{cool}(t) \,.
\end{align}
Here, $F_{\rm cool}$ is a bosonic delta-correlated vacuum noise operator with $\smean{F_{\rm cool}(t)F^\dag_{\rm cool}(t^\prime)}=\delta(t-t^\prime)$ \cite{QuantumNoise}.

\begin{figure}[tb]
\centering
\includegraphics[width=0.41\textwidth]{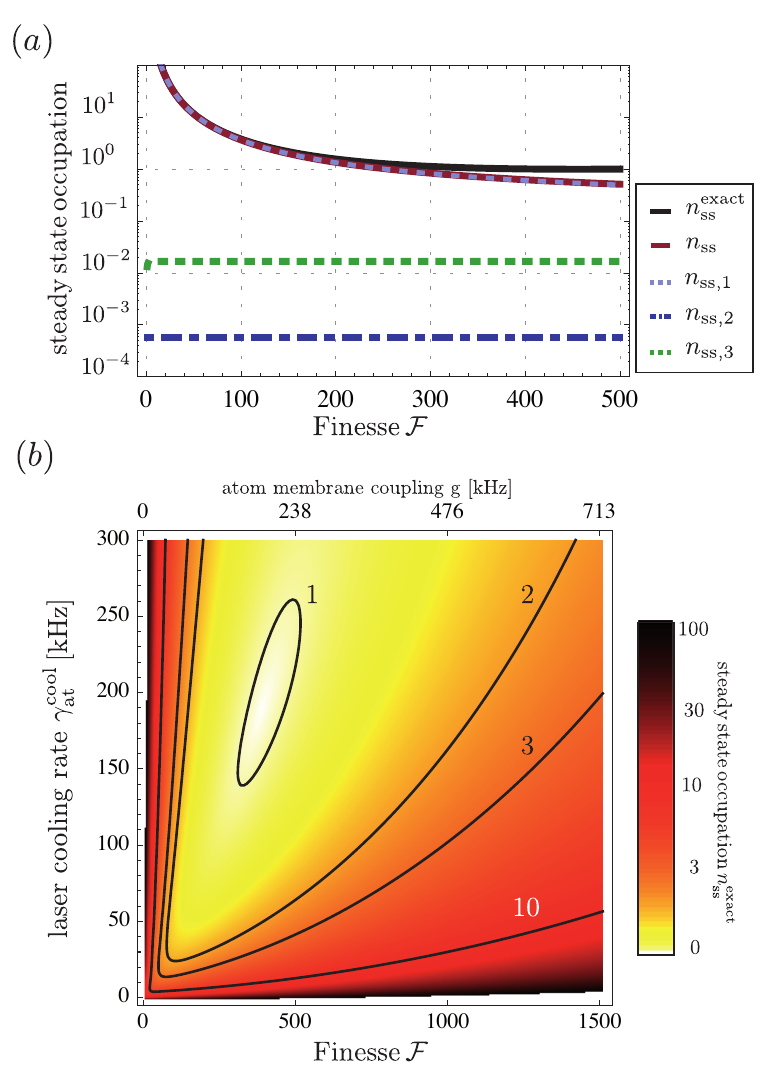}
\caption{Sympathetic cooling of the membrane: Mechanical steady state occupation number.
(a) Approximate total occupation number $n_{\rm ss}$ with its individual contributions in the adiabatic limit (Eq.\,\eqref{eq:occup})  as a function of $\mF$ and for $\gamma_{\rm at}^{\rm cool}=240$ kHz. For comparison, the exact mechanical occupation $n_{\rm ss}^{\rm exact}$, which results from solving the QLEs  \eqref{eomamgscapp} and \eqref{eomaatgsc}, is depicted as a black line. (b) Exact steady state occupation number $n_{\rm ss}^{\rm exact}$ as a function of  $\mF$ and $\gamma_\rmat^{\rm cool}$. The top axis displays the corresponding coherent coupling $g$ and the contours indicate the associated occupation numbers. For both plots the parameters from Table \ref{tab:parameter} have been used.}
\label{fig:coolingfactor}
\end{figure}

To gain insight into the cooling dynamics, we first consider the adiabatic limit $g\ll \gamma_\rmat^{\rm cool}$, where the atomic decay is so fast that the atoms adiabatically follow the membrane. In this case, we may eliminate the atoms and obtain an effective QLE for the mechanical resonator, from which we calculate the EOM for its mean occupation $n=\langle a_\rmm^\dag a_\rmm\rangle$:
\begin{align}
\label{eq:ameff}
\dot{n}(t)=-\left(\frac{\gamma_\rmm}{2}+\frac{\Gamma_{\rm cool}}{2}\right) \left[ n(t) - n_{\rm ss} \right]\,.
\end{align}
Here, we have introduced the sympathetic cooling rate
\begin{align}
\Gamma_{\textnormal{cool}}=\frac{g^2}{\gamma_\rmat^{\textnormal{cool}}}  \frac{1}{1+\left(\gamma_\rmat^{\textnormal{cool}}/4\om_\rmm\right)^2}\,,
\end{align}
as well as the steady state occupation of the mechanical oscillator
\begin{align}
n_{\rm ss}
&\approx \frac{\gamma_\rmm \overline{N}_\rmm+ \gamma_\rmm^{\rm diff}/2}{\gamma_\rmm+\Gamma_{\textnormal{cool}}}+\left(\frac{\gamma_\rmat^{\textnormal{cool}}}{4\om_\rmat}\right)^2 + \frac{\gamma_\rmat^{\rm diff}}{2 \gamma_\rmat^{\rm cool}} \notag \\
&\equiv n_{{\rm ss},1}+n_{{\rm ss},2}+n_{{\rm ss},3}.
\label{eq:occup}
\end{align}
The $n_{{\rm ss},i}$ represent the contributions due to mechanical heating, 'rotating' terms in the coupling ($a_\rmm a_\rmat$, $a_\rmm^\dag a_\rmat^\dag$), and atomic heating, respectively, and we display them in \figref{fig:coolingfactor}(a) as a function of finesse for the parameters of Tab.\,\ref{tab:parameter}. It is clear that the total occupation number is dominated by the mechanical noise $n_{{\rm ss},1}$, which we thus discuss in detail in the following.
Note that we have $\Gamma_{\rm cool}\propto \mF^2$, while $\gamma_\rmm^{\rm diff}\propto\mF^2 $, and the laser-induced contribution to $\overline{N}_\rmm$ scales as $\mF$. For $\gamma_\rmm\ll\Gamma_{\rm cool}$ we thus obtain
\begin{align}
n_{{\rm ss},1}\approx\frac{2 \gamma_{\rm at}^{\rm cool}\omega_L}{N\omega_\rmat^3 m c^2 } P \left( 1 + \frac{a}{\mF}\right) + b\frac{N_\rmm}{\mF^2}\,,
\end{align}
where $a=\pi k_B M c^2 \mathfrak{a}_\rmm^2 \gamma_\rmm / 2 \hbar\omega_L|\mathfrak{r}_\rmm|^2 K_{\rm th}$ and 
$b=\gamma_{\rm at}^{\rm cool}M\pi^2 \gamma_\rmm/4\om_\rmat^2 m |\mfr_\rmm|^2N$.
Here, the three contributions are due to light-induced diffusion, laser-induced heating, and non-zero support temperature, respectively. With the prefactors in the above expression fixed by the considerations of \secref{sec:exprel}, we conclude that 
within the adiabatic limit it is advantageous to work with
high finesses.

While the above argument yields simple analytic expressions that can be used to guide the search for optimal parameters, it is only valid as long as $g\lesssim \gamma_{\rm at}^{\rm cool}$. However, to optimize the parameters we need a solution also valid for larger coupling $g$, which can be obtained by solving the QLEs \eqref{eomamgscapp} and \eqref{eomaatgsc} exactly. For small finesses $\mathcal{F}  < 500$ and fixed laser cooling rate the resulting mechanical occupation number $n_{\rm ss}^{\rm exact}$ is displayed as a black line in  \figref{fig:coolingfactor}(a), showing that the deviations of the two approaches in this parameter range are not too large. For higher finesses, a more complete picture can be gained from  \figref{fig:coolingfactor}(b), where we show the full solution as a function of finesse (coupling) and cooling rate.

Clearly, increasing the finesse beyond a certain optimal value does {\it not} yield lower occupation numbers, in contrast to the tendency derived above. However, the optimal value of the finesse roughly occurs for $g \approx \gamma_{\rm at}^{\rm cool}$, that is on the border of the adiabatic regime.
In particular, we find $n_{\rm ss}^{\textrm exact} \lesssim 1$  for $\mF\approx 450$ and $\gamma_{\rm at}^{\rm cool}\approx 220$ kHz. We thus conclude that ground state cooling of the membrane is possible for realistic experimental parameters.

Besides the formal similarity of our cooling scheme to optomechanical cavity cooling, there are two important differences. First, the atomic dissipation is \emph{tunable}, i.e., after cooling the membrane close to its ground state one can switch off the dissipation and then observe coherent dynamics between the two systems (\secref{sec:cohdyn}). This is not the case in cavity optomechanics, where the cavity decay is fixed. Second, our scheme does not require `resolved sideband' conditions for the cavity. On the contrary, we want to operate in a regime where $\kappa\gg\omega_\rmm,\omega_\rmat$, which greatly relaxes the experimental requirements on the cavity. In this sense, the cooling scheme described here is complementary to standard optomechanical cavity cooling.

\section{Experimental Realization}
\label{sec:exprel}

In order to experimentally realize this proposal, we intend to use a micromechanical membrane in a Fabry-P\'erot cavity in the so-called `membrane-in-the-middle' geometry~\cite{Thompson2008,Jayich2008}. 
We will pre-cool the membrane-cavity setup in a cryostat where high mechanical quality factors up to $10^7$  \cite{Zwickl2008} as well as low absorption of $10^{-6}$ in the near infrared \cite{Sankey2010} have been observed.
In particular, we aim to use a stoichiometric SiN membrane of dimension 1~mm$^2$ with a thickness of $d_{\rmm}=50$~nm.  Such a membrane has an effective mass of $M=3.6\times10^{-11}$~kg and a fundamental mechanical mode near $\om_{\rmm}=2\pi\times400$~kHz.

On the atomic side, we consider using an ultracold gas of $^{87}$Rb atoms. Because the coupling $g$ scales as $\sqrt{N}$, obtaining large atom numbers is crucial to reaching the quantum coherent regime and achieving ground state sympathetic cooling.  Atom numbers as large as 3$\times10^8$ have been prepared in the ground state of a large volume 3D lattice using Raman sideband cooling \cite{Kerman2000}.  We envision using a far detuned two-dimensional (2D) lattice to provide confinement in the transverse direction in addition to the longitudinal lattice that provides the coupling to the membrane. The resulting 3D lattice isolates the atoms from each other, mitigating trap loss due to light-assisted collisions \cite{Kerman2000}. Choosing blue detuning for the coupling lattice reduces spontaneous photon scattering, allowing for reduced laser detunings. This enables lower input powers to achieve the same trap frequency so that larger values of the cavity finesse can be obtained without causing too much absorption-induced heating of the membrane.

Using the expressions for couplings and decoherence processes (including membrane heating), one can find experimental parameters that optimize the ratio of total decoherence to coherent coupling or the ground state membrane occupation.  In doing so, a clear trend towards minimizing the atomic detuning emerges.  However, such a prediction neglects the multi-level structure of the atom.  For a sufficiently large lattice detuning and linear polarization, all ground-state hyperfine $m_F$ levels experience the same trapping potential \cite{Grimm2000}.  At smaller detunings, the potential becomes state-dependent and optical pumping between different $m_F$ levels induced by scattering of photons from the lattice light will reduce the coupling to the membrane.  These effects can be reduced by tailoring the lattice polarization or choosing a laser cooling technique that spin-polarizes the atoms \cite{Kerman2000}.  In order to limit such effects, we consider $\delta/2\pi=1$~GHz as a compromise.  
At the position of the atoms, we set the waist of the beam to $350$ $\mu$m which corresponds to $\mA = 9.6 \times 10^{-8}$ m$^{2}$.
Fixing the value of $\delta$ and $\mA$ fixes the value of the input laser power so as to match the trap frequency to the mechanical frequency of the membrane. This leaves $\mF$ as the remaining experimental parameter, which can be chosen by following the discussion in Secs. \ref{sec:cohdyn} and \ref{sec:sympcool}.
As an example, we provide an optimized parameter set for $\om_\rmm/2\pi=400$~kHz and $\delta/2\pi=1$~GHz in Table~\ref{tab:parameter}. Values for $N$, $Q_\rmm$, and $\mathfrak{a}_\rmm^2$ have been measured previously~\cite{Kerman2000,Zwickl2008,Wilson2009,Sankey2010}.  The value of $K_{\rm th}$ is more uncertain as it is related to the heat conductivity of the membrane, which depends strongly on temperature, material composition and other factors.  We have chosen a value consistent with available literature (see \appref{app:abs}). 

The inhomogeneous intensity profile of the lattice laser beams will lead to some spread $\Delta\om_\rmat$ of atomic vibrational frequencies across the lattice, leading to dephasing of the atomic center-of-mass vibrations. This effect can be neglected as long as $\Delta\om_\rmat \ll \gamma_{\rm at}^{\rm cool}$ while laser cooling is applied, or $\Delta\om_\rmat \ll g$ in the case of coherent dynamics. For the numbers in Table~\ref{tab:parameter}, both conditions correspond  to $\Delta\om_\rmat / \om_\rmat \ll 0.1$.

\begin{table}[htbp]
\begin{center}
\begin{tabular}{|c||c|}
\hline
$\om_\rmm$ & $2 \pi \cdot 400$ kHz  \\
\hline
$M$& $3.6 \cdot 10^{-11}$ kg\\
\hline
$Q_\rmm$&$10^7$\\
\hline
$T_0$&$1.6$ K\\
\hline
$K_{\rm th}$&$4 \cdot 10^{-7}$ W/K\\
\hline
$\mathfrak{a}^2_\rmm$&$10^{-6}$\\
\hline
$\mfr_\rmm$&$0.47$\\
\hline
$\mF$&450\\
\hline
\end{tabular}
\begin{tabular}{|c||c|}
\hline
$\om_\rmat$ & $2 \pi \cdot 400$ kHz  \\
\hline
$\om_\rmL$ & $2 \pi \cdot 384$ THz  \\
\hline
$\delta$& $2 \pi \cdot 1.0$ GHz\\
\hline
$m$&$1.44 \cdot 10^{-25}$ kg\\
\hline
$N$&$10^8$\\
\hline
$\mA$&$9.6 \cdot 10^{-8}$ m$^2$\\
\hline
$\mu$&$1.5 \cdot 10^{-29}$ C m\\
\hline
$P$&$2.8$ mW\\
\hline
\end{tabular}
\begin{tabular}{|c||c|}
\hline
$g$ & $214$ kHz  \\
\hline
$\gamma_\rmm^{\rm diff}$ & $60$ kHz  \\
\hline
$\gamma_\rmat^{\rm diff}$& $8$ kHz\\
\hline
$\gamma_\rmm^{\rm th}$&$73$ kHz\\
\hline
$\delta T$&$4$ K\\
\hline
\end{tabular}
\end{center}
\caption{Optimized parameter set based on the considerations in \secref{sec:discussion} and \secref{sec:exprel}. The two left-most columns show mechanical and  atomic parameters, respectively, while the column on the right displays the resulting coupling and decoherence rates.}
\label{tab:parameter}
\end{table}


\section{Conclusions and Outlook}
\label{sec:conclusions}

In this work, we have discussed a hybrid quantum system, where a micromechanical membrane in a cavity is coupled to the center-of-mass motion of a distant atomic ensemble. Placing a cavity around the membrane enhances the coupling compared to the setup demonstrated in Ref. \cite{Camerer2011}, while preserving the modularity. Apart from observing coherent dynamics, the setup also enables bringing a cryogenically pre-cooled membrane close to its ground state by making use of the tunable atomic dissipation.

In addition to cooling and manipulating micromechanical membranes, the proposed setup could also be used for manipulating other objects, such as levitated dielectric nanospheres \cite{RomeroIsart2010,Chang2010}, or even molecules. Finally, coupling to the internal atomic degrees of freedom \cite{Hammerer2009a, Wallquist2010} instead of the motional ones could open additional possibilities for manipulation and cooling, and might also enable even higher couplings.

\section*{Acknowledgments}
We thank A.~Faber for helpful discussions. This work was supported by the Swiss National Science Foundation through NCCR QSIT and NCCR Nanoscale Science,
the Austrian Science Fund (FWF) through SFB FOQUS, 
the DARPA OLE program,
and by the EU networks AQUTE and MALICIA.
KH acknowledges support through the centre of excellence QUEST, and
MTR is supported by an EU Marie Curie International Incoming Fellowship.

\appendix

\section{Treatment of Suspended End Mirror}
\label{app:mirror}

For the sake of completeness we also treat the standard optomechanical setup consisting of a Fabry-P\'erot cavity with a movable end mirror. The refractive index profile modeling this setup is given by \eeqref{eq:n_of_z} with $n_\rmm=0$, such that the mode functions read
\begin{align}
\label{eq:mirrorMF}
u^\prime_\om(z)&=\frac{1}{2i}\begin{cases}
T'_\om (e^{i k z}-e^{-i k z}) & \text{if }0<z<L  \\
\frac{T'_\om}{T_\om^{'*}} e^{i k z}-e^{-i k z} & \text{if }z>L+d
\end{cases}.
\end{align}
Here, the cavity response is given by $T'_\om=T_\om\vert_{n_m=0}=\mft e^{-ikd} /(1-\mfr e^{2ikL})$, which for $|\mfr|\rightarrow1$ has exactly the Lorentzian resonances given in \eeqref{eq:lorentzian}. For small displacements of the end mirror, the coupling of the latter to the light field can be modeled by the Hamiltonian
\begin{align}
H_{\textnormal{m-f}}^\prime= \frac{\mA}{\mu_0} \hat B^{(-)}(0)\hat B^{(+)}(0) z_\rmm\,,
\label{eq:Hintmf_mirror}
\end{align}
as obtained by evaluating Maxwell's stress tensor for an ideal mirror, which is assumed here for simplicity. 
Linearizing the interaction Hamiltonian given in \eeqref{eq:Hintmf_mirror} analogously to \secref{sec:membranefieldint} provides us with the same expression as in the case of the membrane given in \eeqref{eq:Hmf_lin}, but with a modified coupling constant $g^\prime_\rmm= 2\alpha k_\rmL l_\rmm \mF / \pi^{3/2}$.

\section{Elimination of the Light Field}
\label{app:elimination}

We briefly describe the elimination of the light field to obtain effective EOMs for the atomic and mechanical degrees of freedom. To this end we start from the Heisenberg EOMs for the quadrature operators defined by $x_j=(a_j^\dag+a_j)/\sqrt{2}$ and $p_j=i(a_j^\dag-a_j)/\sqrt{2}$, where $j=\{\rmm, \rmat\}$ as well as $x_\om=(b_\om^\dag+b_\om)/\sqrt{2}$ and $p_\om=i (b^\dag_\om-b_\om)/\sqrt{2}$. From the linearized Hamiltonian in \eeqref{eq:Hlin} we obtain
\label{eq:Heisenbergquad} \begin{align}
 \label{eq:Adotxm}
\dot{x}_\rmm&=\om_\rmm p_\rmm \\
\label{eq:Adotpm}
\dot{p}_\rmm&=-\om_\rmm x_\rmm-2 g_\rmm \! \! \int \!\frac{\textnormal{d}\om}{\sqrt{2\pi}} \, x_\om \\
\label{eq:Adotxat}
\dot{x}_\rmat&=\om_\rmat p_\rmat \\
\label{eq:Adotpat}
\dot{p}_\rmat&=-\om_\rmat x_\rmat -2 g_\rmat \! \! \int \!  \frac{\textnormal{d}\om}{\sqrt{2\pi}} \, x_\om \sin[\Delta_\om\tau+\Delta \varphi_\om]\\
\label{eq:Adotxom}
\dot{x}_\om&=\Delta_\om p_\om \\
\label{eq:Adotpom}
\dot{p}_\om&=-\Delta_\om x_\om -2 g_\rmm x_\rmm \notag \\
& \quad \quad \quad \qquad \! \!-2 g_\rmat x_\rmat\sin[\Delta_\om \tau+\Delta \varphi_\om],
\end{align}
where again $\tau=\overline{z}/c$ is the propagation time between the two systems, $\Delta_\om=\om-\om_\rmL$, and $\Delta \varphi_\om=\varphi_\om-\varphi_{\om_\rmL}$. Solving the Heisenberg \eqsref{eq:Adotxom} and (\ref{eq:Adotpom}) for the field quadratures $x_\om$ yields
\begin{align}
\label{eq:Resultxom}
x_\om(t)&=\cos{[\Delta_\om (t-t_0)]}x_\om(t_0)+ \sin{[\Delta_\om (t-t_0)]} p_\om(t_0) \notag \\[1mm]
& \qquad - 2\! \int_{t_0}^t \!\textnormal{d}s \sin{[\Delta_\om (t-s)]}  \\
& \quad \times\Big[ g_\rmm x_\rmm(s) +g_\rmat x_\rmat(s)\sin[\Delta_\om \tau+\Delta \varphi_\om] \Big],\notag
\end{align}
where the first line contains the initial conditions at some time $t_0$.
Under the `bad cavity' assumption made in the main text, the phase difference $\Delta \varphi_\omega$ varies slowly over the frequency range of interest. Therefore, we may  expand it to first order around the laser frequency according to $\Delta \varphi_{\om\approx \om_\rmL} \approx \Delta_\om/\kappa$, where we have assumed that the laser is resonant with a cavity resonance and that $\sabs{\mfr} \rightarrow 1$. Note that these assumptions are equivalent to the ones we made to obtain the Lorentzian in \eeqref{eq:lorentzian}. In the following we thus approximate
\begin{align}
\Delta_\om \tau+\Delta \varphi_\om \approx \Delta_\om \tilde{\tau}\,,
\label{eq:retardation}
\end{align}
with the modified retardation time $\tilde{\tau}=\tau+1/\kappa$ consisting of the propagation time between the two systems $\tau$ as well as the cavity lifetime $1/\kappa$. 

To proceed, we eliminate the field quadratures $x_\om$ by inserting the solution in \eeqref{eq:Resultxom} into \eeqref{eq:Adotpm}, which yields the following EOM for the mechanics
\begin{align}
\dot{p}_\rmm&=-\om_\rmm x_\rmm(t)+\sqrt{\gamma_\rmm^{\rm diff}}F_\rmm (t)  \\
&\quad - 2 g_\rmm g_\rmat \!\int_{t_0}^t  \! \textnormal{d}s \, x_\rmat(s) \left[ \delta_\theta(t+\tilde{\tau}-s) - \delta_\theta(t-\tilde{\tau}-s) \right]. \notag
\end{align}
Here, $F_\rmm(t)$ is a noise term defined below in \eeqref{eq:noisemech}, and $\delta_\theta (t-s) = \int_{\om_\rmL-\theta}^{\om_\rmL+\theta} \! \frac{\textnormal{d}\om}{2\pi} \, e^{-i \Delta_\om (t-s)}$ is a representation of the $\delta$-function peaked on a time-scale  $\sim1/\theta$. The diffusion rate of the mechanical oscillator is defined as $\gamma_\rmm^{\rm diff}=2 g_\rmm^2$. The EOM for the atomic quadrature operator $p_\rmat$ are obtained analogously by inserting \eeqref{eq:Resultxom} into \eeqref{eq:Adotpat}.

Summarizing these results leads to the following QLEs with retarded interactions:
\begin{align}
\dot{p}_\rmm(t)&=-\om_\rmm x_\rmm(t)+g x_\rmat(t-\tilde{\tau}) + \sqrt{\gamma_\rmm^{\rm diff}} F_\rmm(t) \label{eq:momm} \\
\dot{p}_\rmat(t)&=-\om_\rmat x_\rmat(t)+g x_\rmm(t-\tilde{\tau}) + \sqrt{2}g_\rmat F_\rmat(t) \label{eq:momat},
\end{align}
where $g=2 g_\rmat g_\rmm$ denotes the effective coupling between the mechanical oscillator and the center-of-mass mode of the atoms, and the atomic noise term $F_\rmat(t)$ is given below in \eeqref{eq:noiseat}. In the above equations we can neglect the retardation in the coupling terms since the system operators evolve with rates $\om_\rmat, \om_\rmm$ and $g$ much slower than $\tilde{\tau}^{-1}$ as indicated by the hierarchy of timescales given in \eeqref{eq:HierarchyTimescales}. Up to the atomic noise discussed below we thus recover the EOMs given in \eqref{eq:QLEs} of the main text.

The noise operators appearing in the above equations are Hermitian and given by
\begin{align}
F_\rmm(t)  = &-\!\sqrt{2}  \!\int \!\frac{\rmd\om}{\sqrt{2\pi}}\, \Big\{\cos{[\Delta_\om (t-t_0)]}x_\om(t_0) \notag \\[1mm]
&+ \sin{[\Delta_\om (t-t_0)]} p_\om(t_0) \Big\}\label{eq:noisemech}\\[1mm]
F_\rmat(t) =& -\!\sqrt{2} \! \int \! \frac{\rmd\om}{\sqrt{2\pi}}\,  \sin [\Delta_\om \tilde{\tau}] \Big\{ \cos{[\Delta_\om (t-t_0)]}x_\om(t_0) \notag \\[1mm]
&+ \sin{[\Delta_\om (t-t_0)]} p_\om(t_0) \Big\} \label{eq:noiseat}.
\end{align}
To obtain the corresponding correlation functions, we assume that the field modes $b_\om$ are initially in the vacuum state and therefore obey  $\langle b_\om(t_0) b^\dag_{\om'}(t_0) \rangle= \delta(\om-\om')$.

For the correlation function of the mechanical noise term we obtain $\langle F_\rmm(t)F_\rmm(t') \rangle= \delta_\theta(t-t')$. Note that in the absence of the atoms our setup is exactly the one considered in standard cavity optomechanics, and we should thus find similar decoherence mechanisms. In order to compare the above diffusion rate $\gamma_\rmm^{\rm diff}$ to the optomechanical cooling literature we can express it as $\gamma_\rmm^{\rm diff} \propto \frac{(g_0 \alpha_c)^2}{\kappa}$, where we used that the intra-cavity amplitude is given by $\alpha_c=\alpha/\sqrt{2 \kappa}$ and that the single photon optomechanical coupling can be expressed as $g_0=\frac{\partial \om_c}{\partial z_\rmm}l_\rmm= 2 \sabs{\mfr_\rmm} \om_c  l_\rmm/L$ \cite{Thompson2008}. When taking into account that we assumed the `bad cavity' regime where $\kappa\gg \om_\rmm$ and that we drive the cavity on resonance, we 
obtain the same scaling of the diffusion rate as presented in \cite{WilsonRae2008}, which was recently observed in an experiment \cite{Purdy2012}.

The correlation function for the atomic noise term reads $\langle F_\rmat(t)F_\rmat(t') \rangle=-\frac{1}{4} [ \delta_\theta(t+2\tilde{\tau}-t')+\delta_\theta(t-2\tilde{\tau}-t')-2\delta_\theta(t-t')]$.
On the timescale $\om_\rmat^{-1}$ of the atomic evolution the three $\delta$-kicks in the correlation function average to zero due to our assumption of small retardation times ($\om_\rmat \tilde{\tau} \ll 1$, see \eeqref{eq:HierarchyTimescales}). For this reason the atomic noise operator $F_\rmat(t)$ does not occur in the QLEs in the main text in \eqsref{eq:QLEs}
However, this is an artifact of the 1D treatment, since atoms in a standing wave are well-known to undergo momentum diffusion \cite{Gordon1980}. This effect is introduced in \secref{sec:decoherence}.

\section{Membrane Heating Due to Laser Absorption}
\label{app:abs}
In this section membrane heating due to laser absorption is explicitly modeled. For a circular membrane of diameter $l$, a simple analytical solution can be obtained and the case of the square membrane can then be approximately described by including a `geometric prefactor' of order unity. The results of the analytical model are checked against a finite element method (FEM) simulation.

Heat transport inside the SiN membrane is governed by the heat equation for the temperature $T$:
\begin{equation}
\rho c_p \frac{\partial T}{\partial t} = \kappa_{\rm th} \Delta T + Q_{\rm th}\,,
\end{equation}
where $\kappa_{\rm th}$ is the thermal conductivity, $\rho$ the mass density, $c_p$ the specific heat capacity of the membrane material, $\Delta$ the Laplace operator, and $Q_{\rm th}$ is a source term describing the power dissipated per unit volume. For a 2D temperature distribution inside a thin membrane (thickness $d_\rmm\ll l$) in steady-state ($\partial T / \partial t = 0$), the heat equation reads
\begin{equation}
 \Delta T = \frac{1}{r}\frac{\partial}{\partial r} \left( r\frac{\partial T}{\partial r}\right) + \frac{1}{r^2} \frac{\partial}{\partial \phi}\left( \frac{\partial T}{\partial \phi}\right) = -Q_{\rm th}/\kappa_{\rm th}.
\end{equation}
To obtain a simple analytic solution, the case of a circular membrane  placed in the center of the intra-cavity electric field mode is considered.  For simplicity, we assume that the power $P_{\rm abs}$ is absorbed within a circle of area $\mathcal{A}_\rmm=\pi w_\rmm^2$, where $w_\rmm$ is the waist ($e^{-2}$ radius) of the Gaussian field mode at the position of the membrane. Thus, $Q_{\rm th}=P_{\rm abs}/(\mathcal{A}_\rmm d_\rmm)$.  The absorbed power for a membrane sitting on a slope of the intra-cavity intensity standing wave is given by $P_{\rm abs}=\mathfrak{a}_\rmm^2 2 P_{\rm circ}=\mathfrak{a}_\rmm^2 \frac{ 4\mF}{\pi} P$ with the power absorption coefficient $\mathfrak{a}_\rmm^2=1-\mathfrak{r}_\rmm^2-\mathfrak{t}_\rmm^2$. Here, $P_{\rm circ}$ is the intra-cavity circulating running wave power defined by $P_{\rm circ}=\frac{2 \mF}{\pi} P$ and $P$ is the power of the incident laser. The peak laser power for a standing wave inside the cavity is given by $P_{\rm st}=4 P_{\rm circ}$, such that we obtain half peak power at the position of the membrane.
Making use of the azimuthal symmetry ($\partial T / \partial \phi = 0$), the equation simplifies to
\begin{equation}
\frac{1}{r}\frac{\partial}{\partial r}\left( r\frac{\partial T}{\partial r}\right) = \begin{cases}
 -Q_{\rm th}/\kappa_{\rm th} & \textrm{for  $0\leq r \leq w_\rmm$}, \\
 0 & \textrm{for  $w_\rmm < r \leq l/2$}.
\end{cases}
\end{equation}\\
We solve this equation subject to the boundary condition that the frame is at constant temperature, $T(r=l/2)=T_0$, yielding the temperature distribution
\begin{equation}
T(r) = \begin{cases}
T_0 + \delta T - \frac{Q_{\rm th}}{4\kappa_{\rm th}}r^2 & \textrm{for  $0\leq r \leq w_\rmm$}, \\[1mm]
T_0 + \frac{Q_{\rm th}w_\rmm^2}{2\kappa_{\rm th}} \ln \left( \frac{l}{2r} \right) & \textrm{for  $w_\rmm < r \leq l/2$},
\end{cases}
\end{equation}
where
\begin{equation}
\delta T = T(r=0) - T_0 =  \tfrac{Q_{\rm th}w_\rmm^2}{2\kappa_{\rm th}} \left[ \ln \left( \tfrac{l}{2w_\rmm} \right) + \tfrac{1}{2} \right]
\end{equation}
is the temperature increase of the membrane center compared to the frame. 
The average membrane temperature, obtained by integrating $T(r)$ over the membrane, is
\begin{equation}
T_{\rm avg} = T_0 + \tfrac{Q_{\rm th}w_\rmm^2}{4\kappa_{\rm th}} \left[ 1 - 2\left(\tfrac{l}{w_\rmm}\right)^2 \right].
\end{equation}
The thermal link $K_{\rm th}$ connecting the membrane center to the frame is defined as $K_{\rm th} = P_{\rm abs}/\delta T$. Using $P_{\rm abs}=Q_{\rm th} \pi w_\rmm^2 d_\rmm$, we find
\begin{equation}
K_{\rm th} = \frac{P_{\rm abs}}{\delta T} = \frac{2\pi \kappa_{\rm th} d_\rmm}{ \ln \left( \tfrac{l}{2w_\rmm} \right) + \tfrac{1}{2} }.
\label{eq:thermlink}
\end{equation}

The values for $\delta T$ and $K_{\rm th}$ calculated above for a circular membrane are confirmed by a FEM simulation. Running the FEM simulation instead for a square membrane of side length $l$, keeping all other parameters fixed, we obtain similar results but with a geometric prefactor $f_g$ so that
\begin{equation}
\delta T_\mathrm{square} = f_g \, \delta T \quad \textrm{and} \quad K_{\rm th,\mathrm{square}} = K_{\rm th}/ f_g
\end{equation}
are the corresponding quantities for the center of the square membrane. We find that $f_g$ decreases from $f_g=1.075$ for $w_\rmm/l = 0.3$ to $f_g=1.017$ for $w_\rmm/l = 0.01$. This shows that the model of a circular membrane gives an excellent approximation for the square membrane over the relevant parameter range, well within the errors of other parameters (such as $\kappa_{\rm th}$). Note that in general $w_\rmm/l < 0.3$ to avoid clipping of the beam at the membrane frame. 

As shown in Eq.~\eqref{eq:thermlink}, the thermal link $K_{\rm th}$ depends on the membrane geometry, the beam waist, and the thermal conductivity $\kappa_{\rm th}$.  Because $\kappa_{\rm th}$ depends strongly on temperature, material composition, and other factors its value is not precisely known.  In~\cite{Pierson1999}, a room-temperature value of $\kappa_{\rm th}=25-36~\mathrm{W} \mathrm{K}^{-1} \mathrm{m}^{-1}$ is given for stoichiometric SiN grown by chemical vapor deposition. We extrapolate this value to low temperatures using a temperature dependence as in \cite{Zink2004}.

%

\end{document}